\documentclass[conference]{IEEEtran}

\input{Include/packages}
\def\BibTeX{{\rm B\kern-.05em{\sc i\kern-.025em b}\kern-.08em
    T\kern-.1667em\lower.7ex\hbox{E}\kern-.125emX}}

\fancypagestyle{firstpage}{
  \fancyhf{}
  
  \fancyfoot[C]{\thepage}
}

\definecolor{darkgreen}{RGB}{0,120,0}

\newcommand{\betterparagraph}[1]{\noindent \textbf{#1. }}

\newcommand{\revision}[1]{\textcolor{black}{#1}}

\algnewcommand{\LineComment}[1]{\Statex \(\triangleright\) #1}
\algrenewcommand\algorithmicindent{0.5em}

\newcommand{\squishlist}{
 \begin{list}{$\bullet$}
  { \setlength{\itemsep}{0pt}
     \setlength{\parsep}{3pt}
     \setlength{\topsep}{3pt}
     \setlength{\partopsep}{0pt}
     \setlength{\leftmargin}{1.5em}
     \setlength{\labelwidth}{1em}
     \setlength{\labelsep}{0.5em} } }

\newcommand{\squishlisttwo}{
 \begin{list}{$\bullet$}
  { \setlength{\itemsep}{0pt}
     \setlength{\parsep}{0pt}
    \setlength{\topsep}{0pt}
    \setlength{\partopsep}{0pt}
    \setlength{\leftmargin}{2em}
    \setlength{\labelwidth}{1.5em}
    \setlength{\labelsep}{0.5em} } }

\newcommand{\squishend}{
  \end{list}  }
\title{SCAR: \underline{Sc}heduling Multi-Model \underline{A}I Workloads on Heterogeneous Multi-Chiplet Module Accelerato\underline{r}s \vspace{-1truemm}}

\author{%
\IEEEauthorblockN{Mohanad Odema}
\IEEEauthorblockA{\textit{Univ. of California, Irvine} \\
Irvine, CA, USA \\
modema@uci.edu \vspace{-8truemm}}
\and
\IEEEauthorblockN{Luke Chen}
\IEEEauthorblockA{\textit{Univ. of California, Irvine} \\
Irvine, CA, USA \\
panwangc@uci.edu \vspace{-8truemm}}
\and
\IEEEauthorblockN{Hyoukjun Kwon}
\IEEEauthorblockA{\textit{Univ. of California, Irvine} \\
Irvine, CA, USA \\
hyoukjun.kwon@uci.edu \vspace{-8truemm}}
\and
\IEEEauthorblockN{Mohammad Abdullah Al Faruque}
\IEEEauthorblockA{\textit{Univ. of California, Irvine} \\
Irvine, CA, USA \\
alfaruqu@uci.edu \vspace{-8truemm}}
}

\begin{document}
\maketitle
\thispagestyle{empty}
\pagestyle{empty}

\begin{abstract}

Emerging multi-model workloads with heavy models like recent large language models significantly increased the compute and memory demands on hardware. To address such increasing demands, designing a scalable hardware architecture became a key problem. Among recent solutions, the 2.5D silicon interposer multi-chip module (MCM)-based AI accelerator has been actively explored as a promising scalable solution due to their significant benefits in the low engineering cost and composability. However, previous MCM accelerators are based on homogeneous architectures with fixed dataflow, which encounter major challenges from highly heterogeneous multi-model workloads due to their limited workload adaptivity.

Therefore, in this work, we explore the opportunity in the heterogeneous dataflow MCM AI accelerators. We identify the scheduling of multi-model workload on heterogeneous dataflow MCM AI accelerator is an important and challenging problem due to its significance and scale, which reaches O($10^{56}$) even for a two-model workload on 6x6 chiplets. We develop a set of heuristics to navigate the huge scheduling space and codify them into a scheduler, SCAR, with advanced techniques such as inter-chiplet pipelining. Our evaluation on ten multi-model workload scenarios for datacenter multitenancy and AR/VR use-cases has shown the efficacy of our approach, achieving on average 27.6\% and 29.6\% less energy-delay product (EDP) for the respective applications settings compared to homogeneous baselines. 

\end{abstract}


\section{Introduction}
\label{sec:introduction}

Recent artificial intelligence (AI) inference workloads have increased their scale in both of the model size (e.g., large language models~\cite{brown2020language, touvron2023llama}) and the number of models deployed together (e.g., augmented and virtual reality; AR/VR~\cite{kwon2023xrbench}), which constructs multi-model workloads with heavier models than those in the past. Such trends led to heavy demands on compute capabilities in AI hardware from edge to cloud devices. As an approach to scale up the hardware for AI and increase the compute capability, chiplet-based multi-chip module (MCM) package has emerged as a promising solution~\cite{shao2019simba, tan2021nn, wang2022ai, orenes2023massive}. Such MCM packages facilitate the scaling of AI hardware based on their composability and cost-effectiveness, unlike monolithic designs, which are often constrained by fabrication yields, power, heat, and other engineering costs such as verification~\cite{naffziger2021pioneering}.

Researchers have actively explored the MCM for AI, focusing on the dataflow mapping (i.e., loop ordering, parallelization, and tiling) \revision{of each layer} and workload orchestration onto chiplets considering the network-on-package (NoP) and other communication constraints~\cite{shao2019simba, tan2021nn, wang2022ai, orenes2023massive}. For example, Simba~\cite{shao2019simba} proposed a scalable MCM inference architecture that enables chiplets to either act as standalone inference engines or collaborate as groups for a layer. Although such works have successfully delivered promising performance and energy efficiency than monolithic designs, they mostly focused on \textit{single-model} workloads targeting \textit{homogeneous} chiplets. Unlike single-model workloads, multi-model workloads introduce major challenges to such homogeneous MCMs because of the ML operator heterogeneity (e.g., operator types and tensor sizes) and resulting diverse dataflow preferences~\cite{kwon2021heterogeneous}. Also, multi-model workloads often involve model level dependency and concurrency~\cite{kwon2021heterogeneous, kwon2023xrbench, Sokratis2023multitasc, kim2023dream, panopoulos2024carin}, which adds complex considerations to the scheduling problem.

\begin{figure*}[!tbp]
\begin{center}
{\includegraphics[,width = 0.99\textwidth]{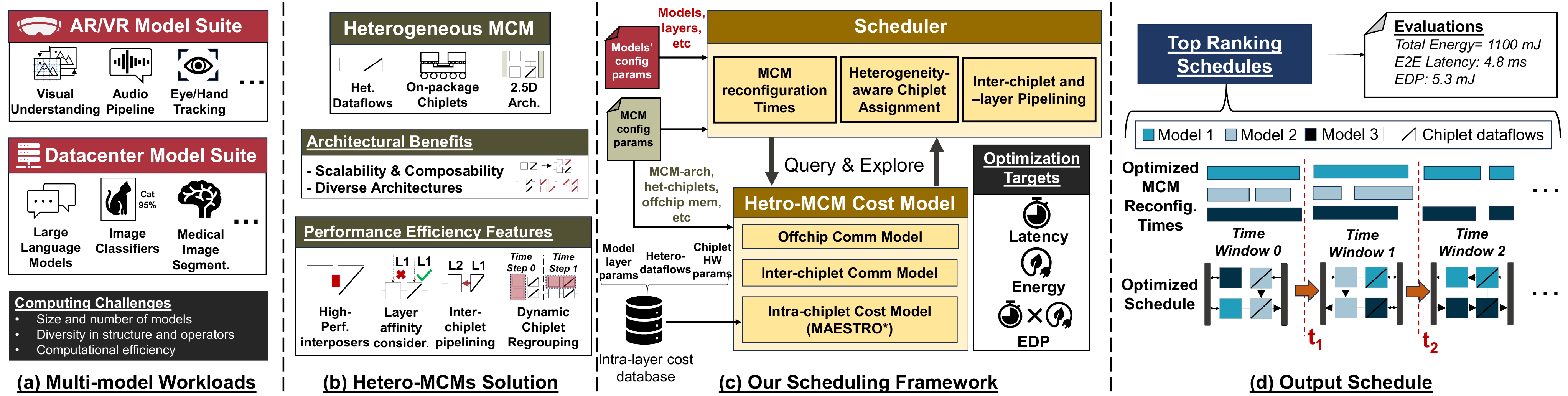}}
\end{center}
\vspace{-2ex}
\caption{An overview of this work: (a) Emerging Multi-model workloads have introduced new challenges for AI hardware. (b) Heterogeneous MCMs present a promising solution to scale with multi-model workloads with some considerations. (c) Our proposed scheduling framework addresses said challenges to explore the heterogeneous scheduling space. (d) Provided Solutions provide optimized spatio-temporal scheduling strategies for the multi-model workloads.
}
\label{fig:Overview}
\vspace{-3ex}
\end{figure*}

Therefore, considering the new trend with multi-model AI workloads \revision{in industry}, such as multi-tenancy~\cite{hazelwood2018applied, wu2019machine, li2022miso} and AR/VR~\cite{kwon2023xrbench}, we explore heterogeneous chiplet-based MCM with AI accelerator chiplets with various dataflows, \revision{as a future-proof option.} \revision{To exploit the benefits of heterogeneous MCM accelerators,} we consider inter-layer pipelining to enhance in-package data reuse and reduce offchip traffic. We formulate the scheduling problem and develop effective heuristics to navigate the huge scheduling space, \revision{whose problem scale is as big as O($10^{56}$) even for a two-model workload (ResNet-50 \cite{he2015deep} and UNet \cite{unet}) on a 6x6 chiplet MCM AI accelerator system (as in Simba \cite{shao2019simba})}.

We evaluate ten MCMs including seven heterogeneous MCMs on ten multi-model scenarios: the first five scenarios are curated using MLPerf inference benchmark \cite{reddi2020mlperf} representing datacenter multi-tenancy scenarios. The models are selected based on recent datacenter model usage trends \cite{hazelwood2018applied, jouppi2017datacenter} and the trend of language model adoptions (e.g., GPT-L~\cite{radford2019language}), future-proofing emerging AI workloads such as AI assistant~\cite{microsoft_copilot}. The other five scenarios are curated for AR/VR usage scenarios from XRBench as a practical use case for edge multi-model workloads ~\cite{kwon2023xrbench}.

The evaluation results show \revision{that heterogeneous MCM combined with our scheduling method is promising for heavy multi-model workloads, which is projected by recent trend.} Compared to the homogeneous MCM~\cite{shao2019simba} running NVDLA~\cite{nvdla} and Shi-diannao~\cite{du2015shidiannao} style dataflows, heterogeneous MCM, on average, achieved 27.6\% and 29.6\% less energy-delay product (EDP) in each domain, respectively. 
\revision{
We also showcase that our scheduler can identify schedules that can reduce EDP to 0.3$\times$ that of single-model schedulers like NN-baton \cite{tan2021nn}.}
We summarize our contributions as follows:

\squishlist
    {\item \revision{As a promising future-proof architecture for heavy multi-model workloads,} we explore heterogeneous dataflow MCM for emerging AI workloads with multiple models running concurrently.}
    {\item We formulate the MCM AI accelerator scheduling problem into a multi-tiered optimization problem to address intractably large scheduling space.}
    {\item Based on the formulation, we develop a scheduler that thoroughly considers heterogeneous MCM and multi-model workloads. The scheduler employs advanced scheduling techniques (inter-layer pipelining, dynamic chiplet regrouping) with resource allocation tree representation~\cite{cai2023inter}.}
    {\item We codify our scheduling method and integrate it with a heterogeneous MCM AI accelerator cost model. We extend MAESTRO~\cite{kwon2019understanding, kwon2020maestro} to model the latency and energy of MCM accelerators.}
    {\item We analyze the costs and benefits of heterogeneous dataflow MCM using industry use case-inspired multi-model workloads and present the importance of the scheduling problem.}
\squishend

\section{Background and Motivation}
\label{sec:background}

\subsection{Multi-model AI Workloads}
\label{subsec:multimodel_workload}

\textbf{Multi-model AI Workloads. }The success of AI algorithms in individual tasks (e.g., hand tracking, depth estimation, speech recognition) led to the emergence of multi-model AI workloads, which include multi-tenant workloads at data centers~\cite{hazelwood2018applied, wu2019machine, li2022miso} and real-time multi-model workloads such as AR/VR~\cite{kwon2023xrbench}. 
We summarize example multi-model AI workloads from industrial use cases in~\autoref{tab:scenarios}.
The models in such workloads are diverse in terms of the tasks and input modalities. For example, an industrial data center multi-tenant AI workload suite~\cite{hazelwood2018applied} includes a face recognition model based on support vector machine, recommendation models based on multi-layer perceptron, and a speech recognition model based on recurrent neural network (RNN).
More recent workloads in data center AI workload include large language models~\cite{azure_openAI}, which adds more heterogeneity to the multi-model AI workloads.
As discussed in prior works~\cite{kwon2021heterogeneous, kwon2023xrbench}, such multi-model workloads involve high heterogeneity in AI operators (or layers), which is one of the major challenges to accelerators that specialize the architecture and dataflow for a specific set of workloads.

\subsection{Scheduling AI workloads on AI Hardware and MCMs}
\label{subsec:bg_mcm_accelerators}

Scheduling AI workloads considers the assignment of computations (e.g., model, layer, or tile) to target hardware platforms and their constituent computing units. We provide a brief on the state of AI workloads scheduling practices. 

\textbf{Scheduling on CPU/GPU systems.} modern systems (servers) typically employ GPUs and/or CPUs for inference services \cite{hazelwood2018applied, choi2022serving, crankshaw2017clipper, hauswald2015djinn, olston2017tensorflow, shen2019nexus, wu2019machine, zhang2019mark, baek2020multi}. Most of these computing units are based on homogeneous cores -- or simple heterogeneity such as big and little cores in CPUs, or CUDA and Tensor cores in GPUs. Traditionally, the scheduling in such settings is concerned with the coarse assignment of models to computing units, leaving the operator assignments to be performed in a direct manner (e.g., all GEMM operations to Tensor Cores in a GPU). 
As multi-model workloads proliferated, new features (e.g., GPU sharing) emerged to 
improve inference services for small-batch inference tasks \cite{choi2022serving, choi2021multi}. Still, the limited programmer/compiler control and the cache-based memory systems restrict CPUs/GPUs from engaging multi-model workload scheduling on a finer granularity. 

\textbf{Scheduling on customized AI accelerators.} Customized AI accelerators (Google's TPU \cite{jouppi2017datacenter} Meta's MTIA \cite{firoozshahian2023mtia}) enable full programmer/compiler control over memory operations (when and what to read/write, when and what to evict, etc). AI accelerators typically employ scratchpad memory-based systems to support deterministic low-level activities \cite{ghodrati2020planaria, kwon2020maestro, parashar2019timeloop, kwon2021heterogeneous}. AI accelerators are also integrated into edge hardware (e.g., NPU in Apple Vision Pro's M2 chip \cite{apple_npu}).

\textbf{Scheduling on MCM AI Accelerators. }To scale with the rising compute demands of modern AI workloads, multi-chip modules (MCMs) have emerged as viable approach enabling the integration of composable, small functional dies (chiplets) on the package level to build a larger system, where they are connected together via on-package links typically through silicon interposer or organic substrates to create a network-on-package (NoP) \cite{vivet20202, kannan2015enabling, beck2018zeppelin}. Through enabling scalability via adjusting the number of chiplets on the package, as well as low verification costs \cite{naffziger2021pioneering}, many chiplet-based systems have been developed for scalable deep learning inference: Simba \cite{shao2019simba} is one comprising 36 chiplets, each containing 16 processing engines to deliver up to 128 TOPs computing capability. 
Another example is Tesla's DOJO chiplet-based architecture capable of scaling to exaFLOP supercomputers for large-scale machine learning.  
The scaling in chiplet sizes, architectures, and computational capabilities 
has enabled support for serving multi-model workloads together on the same MCM system with a finer degree of scheduling granularity (operator, tiles). 
However, multi-model schedulers face new challenges compared to their single-model counterparts considering the increased memory footprints, bandwidth contention, etc. 

\subsection{Motivational Example}
\label{subsec:challenges}

\begin{figure}[!tbp]
\begin{center}
{\includegraphics[,width = 0.46\textwidth]{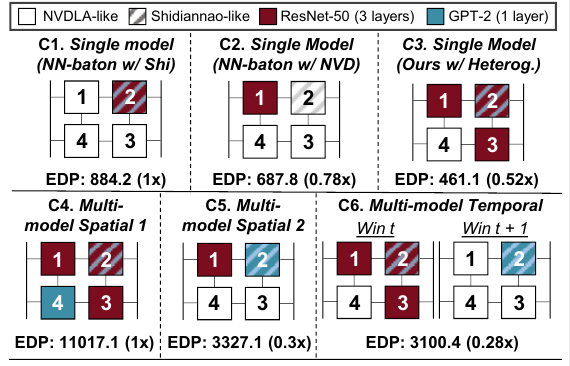}}
\end{center}
\vspace{-2.2ex}
\caption{Motivational study on a $2\times2$ MCM AI accelerator using batch size of 1 for
3 layers from the second ResNet-50 block and the first feed forward layer from GPT-2. Each chiplet has 4096 PEs and 10 MB L2 shared memory (Full from Section \ref{sec:eval}). NN-baton \cite{tan2021nn} considers partitioning computation across chiplets only when not enough resources exist.
}
\label{fig:sngl_multi}
\vspace{-3.8ex}
\end{figure}

We consider the NN-baton \cite{tan2021nn} as our baseline scheduler as it targets scheduling single model workloads on multi-chiplet accelerators. NN-baton proposes to partition a single model workload across several chiplets whenever its computational demands exceeds a single chiplet's capacity, and employs a unified dataflow across the chiplets.
As heterogeneous accelerators proliferate \cite{kwon2021heterogeneous, zhang2022h2h}, chiplets technology has facilitated their integration on the package level. 
Consider a small heterogeneous 2$\times$2 MCM containing 3 NVDLA-like (weight stationary) and 1 Shidiannao-like (output stationary accelerators, and consider a small multi-model workload constituting 3 layers from the second ResNet-50 block and one fully connected layer from GPT-L. We analyze the schedules yielded through NN-baton and our scheduler as follows.

\textbf{Single model case.} We show the single model scheduling results for the ResNet-50 workload in Figure \ref{fig:sngl_multi} [A1-A3]. As each chiplet possess sufficient resources to process the ResNet-50 workload, NN-baton schedules the workload onto a single chiplet. As shown, scheduling the ResNet-50 workload to the NVDLA-like chiplet (A2) experiences 0.78$\times$ the EDP as that from the Shidiannao-like chiplet (A1). However, a more nuanced schedule (A3) identified through our scheduler leverages heterogeneity by distributing the ResNet-50 layers across the heterogeneous chiplets, sustaining 0.52$\times$ less EDP than (A1) through catering to individual layer affinities.

\textbf{Multi-model case.}, In Figure \ref{fig:sngl_multi} (B1-B3), NN-baton (B1) is agnostic to the heterogeneous MCM composition, executing each model workload sequentially on its starting chiplet 1. We show two schedules that are sampled through our schedules: (i) In (B2), our scheduler can spatially distribute the ResNet-50 and GPT-2 workloads across the NVDLA-like and Shidiannao-like accelerators, respectively, leading the EDP to be 0.3$\times$ that in (B1). (ii) In (B3), Our scheduler recognizes a spatio-temporal optimization to leverage the heterogeneous chiplet pipelining for the ResNet-50 in one time window, and then schedule the GPT-2 layer on the Shidiannao layer in the following time window, leading EDP to be 0.28$\times$ that of (B1).

\subsection{Complexity of the multi-model scheduling space}
\label{subsec:complexity_challenges}

To understand the scale of the multi-model scheduling problem, we analyze its search space complexity.
Let a multi-model workload constitute $N$ models, each model containing $L_i$ layers, and $L$=$\sum_i^N L_i$. Let $C$ be the total number of accelerator chiplets on an MCM. Then, a characterization of the multi-model scheduling space can be given as $\mathcal{O}(C^L \times \frac{L}{L_1!L_2!\cdots L_N!})$. 

The first term ($C^L$) covers the set of possible chiplet assignments for each layer (spatial complexity); whereas the multinomial coefficient ($\frac{L}{L_1!L_2!\cdots L_N!})$ covers the number of ways to interleave multiple sequences of layers, while maintaining the layer dependencies for each model. In the small motivational example above, this complexity accounts for a total of $\mathcal{O}(1536)$ scheduling possibilities. If we consider a more practical case involving a ResNet-50 and UNet models ($L1=50$ and $L2=23$) on a full Simba system ($C=36$), the complexity becomes $\sim$ $\mathcal{O}(10^{56})$, showcasing an exponential rise as the models grow in number and complexity. 

\subsection{Summary on the challenges of multi-model scheduling}
\label{subsec:scheduling_challenges}

We summarize the unique scheduling challenges for multi-model workloads compared to their single-model counterparts:

\begin{itemize}
\item \textbf{Layer sequence permutations. }
In a single model scenario, the space of computation assignments is defined based on dependent layer sequences from a single model. Whereas in a multi-model scenario, independent layer sequences from different models also exist, compounding the decision space complexity as a result of permutations.


\item \textbf{Spatial mapping conditioning. }
The quality of one model's schedule is affected by the spatial mappings of other models' mappings due to resource availability, bandwidth contention, added data travel times, and so on.

\item \textbf{Heterogeneous Integration Trade-offs. }
Performance efficiency depends on the underlying pattern of heterogeneous integration, and the diversity within and across model workloads imply that no single pattern fits all.

\end{itemize}

\begin{figure*}[!tbp]
\begin{center}
{\includegraphics[,width = 0.99\textwidth]{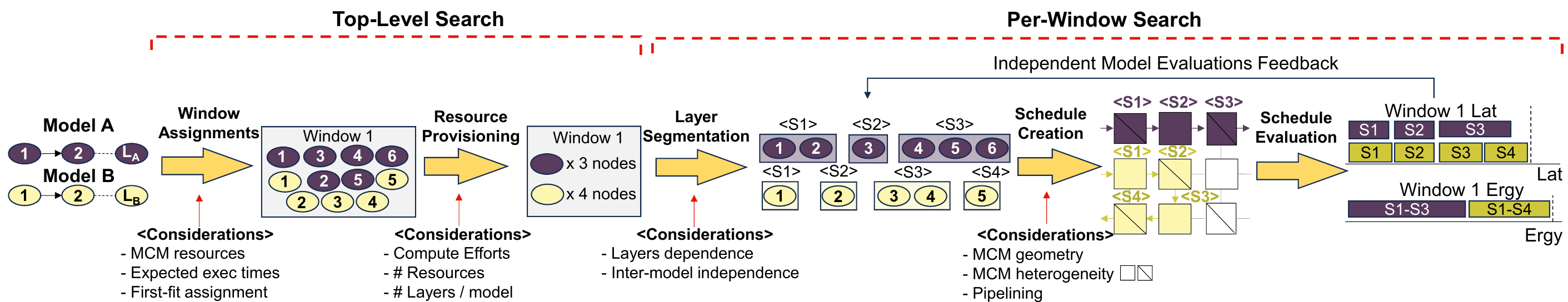}}
\end{center}
\vspace{-2ex}
    \caption{An overview of our scheduling algorithm for multi-model workloads on heterogeneous MCM AI accelerators. }
\label{fig:algorithm}
\vspace{-2ex}
\end{figure*}

To address the challenges and search complexity, one approach is to formulate the problem as a multi-level decision problem where each decision subspace is a tractable problem~\cite{kwon2021heterogeneous,chatarasi2021marvel}. We adopt a similar approach and formulate the MCM multi-model workload scheduling as multiple-level decision problem, as shown in Figure \ref{fig:algorithm}. We detail our problem formulation and performance modeling methodology next.
\section{System Modeling and Problem Formulation}
\label{sec:problem_formulation}

To develop a systematic approach to navigate complex search space, we formulate the scheduling problem of multi-model workloads on a heterogeneous MCM AI accelerator.

\subsection{Base Formulation}
\label{subsec:formulation_base}

\begin{table}[t]
    \centering
    \caption{Notation used in the formulation.}
    \vspace{-2ex}
    \begin{tabular}{l | l}
    \hline

    \textbf{Notation} & \textbf{Description} \\
    \hline
    $Sc$ & Multi-model workload scenario \\
    $m_i$ & the i-th model from scenario $Sc$ \\
    $layer_{i,j}$ & the j-th layer from model i \\
    \hline
    $H$ & MCM hardware \\
    $C$ & Set of accelerator chiplets on $H$ \\
    $c_i$ & the i-th chiplet from $C$ \\
    $DF$ & set of supported dataflows on $H$ \\
    $n_{df_i}$ & Number of chiplets adopting the i-th dataflow \\
    $BW_{offchip}$ & offchip bandwidth \\
    $BW_{nop}$ & Network-on-package bandwidth \\
    \hline
    $df$ & Dataflow \\
    $N_{PE}$ & Number of processing engines \\
    $BW_{noc}$ & Network-on-chip bandwidth \\
    $Sz_{mem}$ & The memory size in $c$ \\
    \hline
    $TW(Sc)$ & The set of time windows for $Sc$ \\
    $tw(Sc)$ & an execution time window for $Sc$ on $H$ \\
    $T_s$ & time window start \\
    $T_{tw}$ & time window end \\
    $L(tw(Sc))$ & set of layers executable on $H$ during $tw(Sc)$ \\
    $SG$ & Set of all valid segments for $tw(Sc)$ \\
    $sg(tw(Sc))$ & a layer segment from $L(tw(Sc))$ \\
    \hline
    $Sp_{tw}(Sc)$ & Time window partioning space for $Sc$ \\
    $Sp_{sg}(tw)$ & Layer segmentation space at $tw$ \\
    $SS_{TW}(tw(Sc), H)$ & The scheduling space for $tw(Sc)$ on $H$ \\
    $SS_{Sc}(H)$ & The overall scheduling space for $Sc$ on $H$ \\
    $sched(Sc, H)$ & A scheduling instance for $Sc$ on $H$ \\
    \hline
    $Lat^{i}_j(\mathbb{A})$ & Latency evaluation for $\mathbb{A}$ given identifiers $i,j$ \\
    $E^{i}_j(\mathbb{A})$ & Energy evaluation for $\mathbb{A}$ given identifiers $i,j$ \\
    $Sz_{data}$ & Size of transmission data \\
    $n_{hops}$ & Number of hops from src to destination\\
    $n_{splits}$ & Number of time window splits \\
    \hline

    \end{tabular}
    \label{tab:notation}
    \vspace{-2ex}
\end{table}

To formulate the MCM scehduling problem, we first define multi-model workload scenario ($Sc$) and MCM hardware ($H$).

We formulate the workload in the granularity of layers in each model. Therefore, we formulate a multi-model workload scenario ($Sc$) as the collection of layers in the models included in the scenario. Letting the number of models included in $Sc$ as $|Sc|$ and the number of layers included in a model $m$ as $|m|$, we define $Sc$ as follows: 
\begin{definition} \textbf{Multi-model Workload Scenario (Sc)}
$$Sc = \{layer_{i,j}| 0 < i \leq |Sc|, 0< j \leq |m_i| \}
$$
where $layer_{(i,j)}$ refers to the j-th layer of model i in Sc.
\label{def:workload_scenario}
\end{definition}

AI accelerator chiplets consist of a PE array, memory, and on-chip interconnection among memory and PEs. In addition to them, we also include the dataflow in the formulation to model heterogeneous chiplet MCM AI accelerator. Accordingly, we define an AI accelerator chiplet ($c$) as follows:
\begin{definition} \textbf{AI Accelerator Chiplet (c)}
$$
c = \{df, N_{PE}, BW_{noc}, BW_{mem}, Sz_{mem}\}  
$$
\label{def:chiplet}
\vspace{-7mm}
\end{definition}
In~\autoref{def:chiplet}, $df$ refers to the dataflow, $N_{PE}$ is the number of PEs, $BW_{noc}$ is the NoC bandwidth, $BW_{mem}$ is the chiplet-level shared memory bandwidth, and $Sz_{mem}$ is the memory size in $c$. 

Based on the definition of the chiplet, we formulate the MCM accelerator as the set of chiplets ($C = \{c_1, c_2, ..., c_{N_{cpl}}\}$), NoP, and off-chip interface as follows:
\begin{definition} \textbf{MCM AI Accelerator (H)}
$$H = \{C, BW_{offchip}, BW_{nop}\} $$ 
\label{def:mcm}
\vspace{-7mm}
\end{definition}
Unless otherwise stated, we assume the 2D mesh topology for NoP like Simba~\cite{shao2019simba}, and chiplets on two sides (left and right) of the packages have off-chip interfaces.

\subsection{Workload Partitioning Space}
\label{subsec:formluation_partitioning}

To reduce the complexity of the scheduling problem, we adopt a multi-level scheduling method, which splits the end-to-end workload defined in the layer granularity into coarse-grained layer groups, termed as the \textit{time window}. ~\autoref{fig:algorithm} shows an example of the time window that contains six layers from Model A and five layers from Model B.

A time window ($tw$) is defined by the start time and the duration ($T_{S}$ and $T_{tw}$) and a set of assigned layers to the time window, as shown in~\autoref{def:time_window}.

\begin{definition} \textbf{Time Window (tw)} \\
\noindent
For a target workload scenario $Sc$, a time window $tw$ is defined as follows: 
$$
tw(Sc) = (T_{s}, T_{tw}, L)
$$
where $L = \{l | l \in Sc \}$
\label{def:time_window}
\end{definition}

The time window describes a set of layers to be executed on an MCM AI accelerator package, which is used for describing package level scheduling. For each chiplet, we define a finer-grained group of layers within a time window. We term the sub-set of layers within a time window as \textit{segment}.

\begin{definition} \textbf{Segment (sg)} \\
\noindent
For a time window $tw(Sc)$ and its layers $L(tw(Sc))$, the segment $sg(tw(Sc))$ is defined as follows: 
$$
sg(tw(Sc)) = \{l | l \in L(tw(Sc)) \}
$$
\label{def:segment}
\vspace{-5mm}
\end{definition}

To develop a systematic optimization algorithm for layer segmentation i each time window, we need to define the conditions of valid layer segments, provided  as follows:

\begin{theorem} \textbf{The validity of segments in a time window} \\
\noindent
For a time window $tw(Sc)$ and its layers $L(tw(Sc))$, 
let the set of all segments for $tw(Sc)$ be $SG$, then $SG$ is valid if the following condition is satisfied:
$$
\bigcup_{sg\in SG} sg = L(tw(Sc)) \land [\forall sg_i \neq sg_j \in SG, sg_i \cap sg_j = \emptyset]
$$
\label{thm:segment_correctness}
\end{theorem}
\autoref{thm:segment_correctness} states two conditions (1) the set of segments needs to cover all the layers in their time window for completing assigned layer computations for the time window and (2) all segments are exclusive to prevent redundant computing. The same idea extends to the time window as follows:

\begin{theorem} \textbf{The validity of time window partitioning} \\
For a multi-model workload $Sc$, its layers $L(Sc)$, and the set of time windows $TW(Sc)$, $TW(Sc)$ is valid if the following condition is satisfied:
$$
\bigcup_{tw\in TW(Sc)} tw \text{=} L(Sc) \land [\forall tw_i \neq tw_j \in TW(Sc), tw_i \cap tw_j = \emptyset]
$$
\label{thm:tw_partitioning_correctness}
\end{theorem}

Both \autoref{thm:segment_correctness} and ~\autoref{thm:tw_partitioning_correctness} indicate that the segments and time windows need to be partitions of the workload and time window layers, respectively. Combining all definitions in this section, we formulate the workload partitioning space into the time window and segment as follows:

\begin{definition} \textbf{Workload Partitioning Space} \\
\noindent
For a multi-model workload $Sc$, the time window partitioning space ($Sp_{tw}(Sc)$) and the layer segmentation space for a time window ($Sp_{sg}(tw)$) are defined as follows: 
\begin{align*}
Sp_{tw}(Sc) &= \mathbb{P}(L(Sc)) \\ 
Sp_{sg}(tw) &= \mathbb{P}(L(tw)) 
\end{align*}
where $\mathbb{P}(A)$ refers to all possible partitioning of a set $A$
\label{def:part_space}
\end{definition}

\subsection{Scheduling Space}
\label{subsec:formluation_scheduling}

A segment contains layers to be executed on a chiplet. Therefore, spatial (i.e., which segment runs on which chiplet) and temporal mappings (i.e., execution order of segments on each chiplet) of segments construct the scheduling space within each time window when segments are determined. Therefore, the scheduling space within a time window $tw(Sc)$ can be defined as follows:

\begin{definition} \textbf{Scheduling Space in a Time Window ($SS_{TW}$)} \\
\noindent
For a given time window $tw(Sc)$ and a target MCM accelerator hardware $H$, the scheduling space within the time window ($SS_{TW}(tw(Sc), H)$ is defined as follows:
\begin{align*}
&SS_{TW}(tw(Sc), SG, H) \\
& \text{=} \{(sg, c, j) | sg \in SG \land c \in C_H \land j \in \mathbb{N} \land val(sg, tw(Sc)) \}
\end{align*}
where $C_H$ refers to the set of chiplets in $H$ and val($sg$, $tw(Sc)$) indicates the validity of $sg$ for $tw(Sc)$
\label{def:scheduling_space_time_window}
\end{definition}

Each entry in $SS_{TW}$ describes the spatial and temporal mapping of a segment ($sg$). Spatial mapping can be defined as the target chiplet to execute $sg$. Accordingly, a target chiplet ($c$) is specified for $sg$. The temporal mapping is defined as the execution order. Therefore, a natural number $j$ is used to represent the execution order. Note that the execution order is defined separately on each chiplet. Based on~\autoref{def:scheduling_space_time_window}, we can define the entire scheduling space as the collection of that in each time window.

\begin{definition} \textbf{MCM Scheduling Space for a Multi-model Workload ($SS_{Sc}(H)$)} \\
\noindent
For an MCM AI accelerator ($H$) and a multi-model workload ($Sc$), the scheduling space ($SS_{Sc}(H)$) is defined as follows:
\begin{align*}
SS_{Sc}(H) = &\{(TW, SG_{TW}, (SS_{TW}(tw,SG_{TW}(tw),H)) | \\ 
& TW \subset Sp_{tw}(Sc) \land SG_{TW}(tw) \subset Sp_{sg}(tw) \\
& \land tw \in TW \}
\end{align*}
where $SG_{TW}$ refers to the set of layer segments for each time window in $TW$
\label{def:scheduling_space_full}
\end{definition}

\autoref{def:scheduling_space_full} defines the entire scheduling space of an MCM AI accelerator for a multi-model workload as the cross-product of all possible time window partitioning, layer segmentation for each time window, and corresponding scheduling space within each time window.

\subsection{Scheduling Problem}
\label{subsec:formluation_scheduling_problem}

Based on~\autoref{def:scheduling_space_full}, we define a schedule instance as the collection of spatial and temporal mapping for given valid time windows ($TW$) and segments for each time window ($SG_{TW}$).

\begin{definition} \textbf{MCM Schedule}\\
\noindent
A schedule instance ($sched(Sc,H)$) is defined as follows:
\begin{align*}    
sched(Sc, TW, &SG_{TW}, H) \text{=} \{(TW, SG_{TW}, s) | valid(TW, Sc) \\
&\land \forall tw \in TW: val(SG_{TW}(tw), tw) \\ 
&\land s \in SS_{TW}(tw, SG_{TW}(tw), H)\}
\end{align*}
where $SG_{TW}$ refers to the set of layer segments for each time window in $TW$
\label{def:mcm_schedule}
\end{definition}

Using~\autoref{def:mcm_schedule}, we formulate the scheduling problem as a minimization problem of an optimization metric of choice (e.g., latency and energy), as follows:

\begin{definition} \textbf{MCM Scheduling Problem}
$$
\operatorname*{argmin}_{TW, SG_{TW}, Sched} OptMetric(TW, SG_{TW}, Sched, H) $$
where $Sched = sched(Sc, TW, SG_{TW}, H)$
\end{definition}

The optimization metric can be chosen by users depending on the use case. In our scheduler, we adopt a comprehensive and customizable score that thoroughly considers all of latency, energy, and energy-delay product (EDP),
allowing users to configure their own optimization metrics, which can be the mentioned frequently used metrics, or a user-defined function that takes a schedule instance and generates a custom metric.

\subsection{Performance Modeling}
\label{subsec:lat}

In order to evaluate schedules on target MCM AI accelerator hardware, 
we extend MAESTRO~\cite{kwon2019understanding} to the chiplet domain and model the latency experienced on MCM AI accelerators when concurrently executing multi-model workloads 
We discuss our latency evaluation methodology in detail next.

\betterparagraph{Layer Latency} The latency incurred by an individual layer, $l$, mapped  onto an accelerator chiplet is defined as:
\begin{equation*}
    Lat(l) = Lat^{ip\_com}(l) + Lat^{comp}(l) + Lat^{op\_com}(l)\label{def:lyr_lat}
\end{equation*}
$Lat^{comp}(l)$ is the layer computation cost dependent on chiplet parameters in \autoref{def:chiplet}; $Lat^{ip\_com}(l)$ is latency incurred from loading the layer operands (input activations and weights); $Lat^{op\_com}(l)$ is transmission latency of output activation to a subsequent layer. $Lat^{com}$ is defined as:
\begin{equation*}
    \small
    Lat^{com} = 
    \begin{cases}
        0, \text{if same chiplet} \\
        \frac{Sz_{data}}{BW_{nop}} + n_{hops}\times Lat_{hop} + \delta,  \text{if same package} \\
        \frac{Sz_{data}}{BW_{mem}} + n_{hops}\; \text{x} \;Lat_{hop} + Lat_{mem} + \delta, \text{if offhcip}
    \end{cases}
\end{equation*}
where communication costs are incurred when transmitting data to/from another chiplet on package or the offchip memory. The first term $\frac{Sz_{data}}{BW}$ reflects transmission latency; the second term is captures propagation latency across $n_{hops}$ between the source and destination; $\delta$ is an additional latency term for potential NoP traffic conflicts; $Lat_{mem}$ is the offchip memory read/write latency based on memory bandwidth.

\betterparagraph{Time Window Latency} We first model a layer segment's latency in a time window as follows:
\begin{equation*}
    Lat(sg) = \sum_{n=1}^{N} Lat^{comp}(l_n) + Lat^{ip\_com}(sg) + Lat^{op\_com}(sg)
\end{equation*}
The first term represents the sum of individual layer computational latencies; $Lat^{ip\_com}$ is the initial external off-chiplet data transfer to load inputs and weights; $Lat^{op\_com}$ is the transmission latency from transmitting segment output data to the next segment or offchip memory. From here, we can define the time window latency as:

\begin{equation*}
    Lat(tw) = \max\limits_{SG_m \subset SG}\;\; [Lat(SG_m)]
\end{equation*}
where for a time window's assigned set of segments, $SG$, the overall window latency is the maximum latency incurred by a subset of segments, $SG_m \in SG$, associated with a model $m$ and necessitating sequential execution. $Lat(SG_m)$ can be estimated as the pipelining latency across segments -- that is, given model $m$ requires processing data with batch size, $b$, and the max number of samples any chiplet can process at a time is given by mini-batch size, $b' <= b$, we get:
\begin{equation*}
    Lat(SG_m) \text{=} \sum_{sg_k \in SG_m} Lat(sg_k | b') + (\frac{b}{b'} - 1) \times \max\limits_{sg_k}  [Lat(sg_k | b')]
    \label{eqn:pipeline}
\end{equation*}


\betterparagraph{Overall Latency}The overall Scenario latency can then be estimated as the aggregate across all time windows:
\begin{equation*}
    Lat(Sc) = \sum_{tw_j \in TW} Lat(tw_j)
\end{equation*}

\betterparagraph{Energy Modeling} Albeit similar to latency, energy costs are always aggregated considering communication energy costs of data sizes, $n_{hops}$, bit transmission energy, and memory access. 




\section{Proposed Scheduling Framework: SCAR}
\label{sec:scheduling_framework}

We discuss our scheduling framework, SCAR, for multi-model workloads on heterogeneous MCMs based on the hierarchical search space characterization and problem formulation in 
\autoref{sec:problem_formulation}.
As illustrated in~\autoref{fig:algorithm}, our scheduling algorithm is a two-level approach: top-level and per-window searches. Top-level search is responsible for selecting layers in each model to be scheduled within a time window and determining the initial number of chiplet nodes for each model. Per-window search explores the spatial and temporal partitioning (tiles or layer segments) of the layers in each model at the chiplet granularity. To explore the chiplet granularity tiling space, we generate valid inter-chiplet-pipelined schedules utilizing a scheduling tree structure inspired by the RA Tree~\cite{cai2023inter}. Each schedule is evaluated using our custom heterogeneous MCM cost model which provides feedback to the chiplet level tiling ("layer segmentation" in ~\autoref{fig:algorithm}) with expected metrics (latency, energy, EDP, etc.). 

\begin{figure}[!tbp]
\begin{center}
{\includegraphics[,width = 0.49\textwidth]{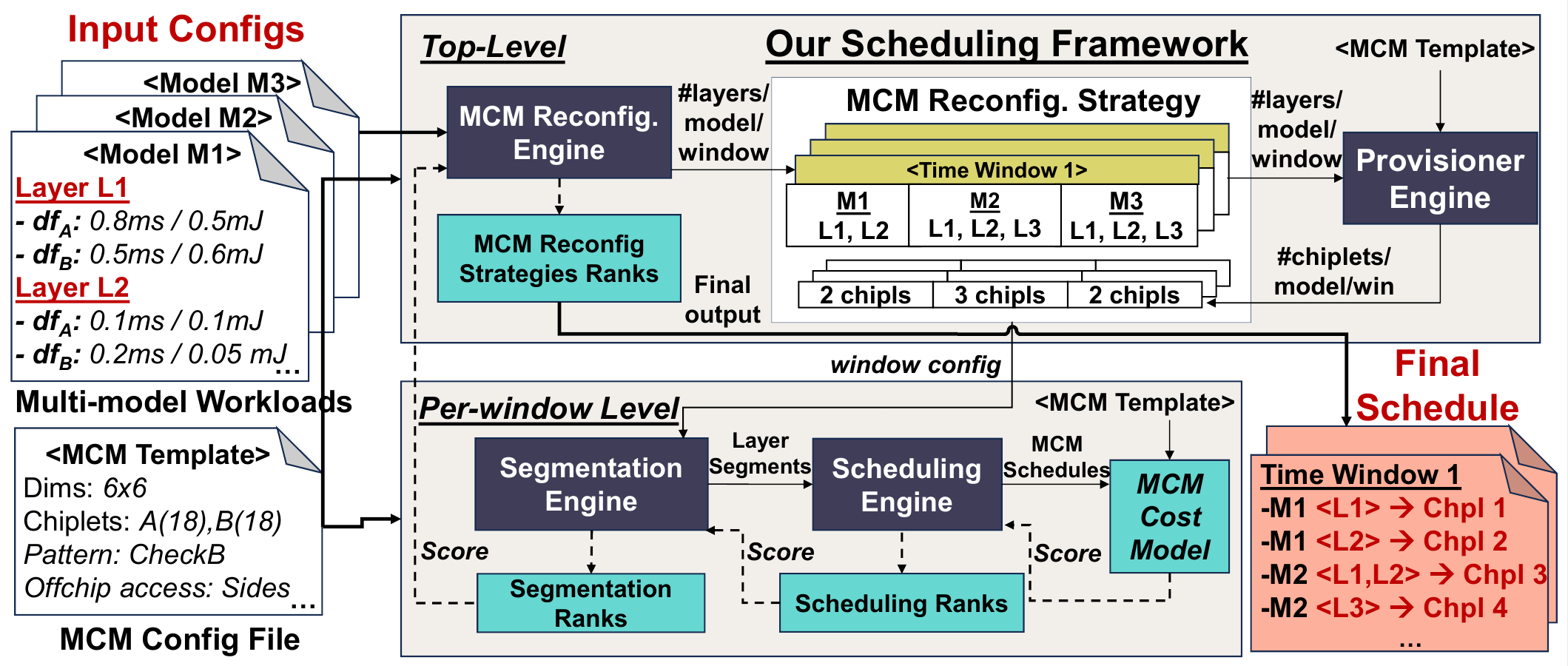}}
\end{center}
\vspace{-2.6ex}
    \caption{Our proposed multi-model scheduling framework on heterog. MCM.}
\label{fig:framework}
\vspace{-2ex}
\end{figure}

We codify our scheduling algorithm into a software framework, as illustrated in~\autoref{fig:framework}. As inputs, our scheduling framework receives (1) description files of the multi-model workloads (layer parameters, topology, dependencies, etc.) and (2) a description file of the MCM hardware specification (the number of chiplets, the shape, and chiplet arrays dataflow organization, NoP bandwidth, on-chiplet memory size, etc.). As outputs, our scheduling framework reports an optimized schedule with expected metrics such as latency, energy, EDP, or other user-defined metrics as a combination of latency and energy. Our scheduling framework consists of four software \textit{engines} illustrated in~\autoref{fig:algorithm} discussed next.

\subsection{MCM Reconfiguration Engine \textit{(MCM-Reconfig)}}
\label{subsec:MCM-reconfig}

The \textit{MCM-Reconfig} engine at the top-level step receives the multi-model workload descriptions with layer information in each model, layer dependency, and expected latency and energy of each layer on each chiplet class offline-analyzed by MAESTRO~\cite{kwon2019understanding}. The \textit{MCM-Reconfig} engine is responsible for the window assignment in~\autoref{fig:algorithm}, which (1) generates candidate time window partitioning strategies via sampling a set of discrete points in time as boundary points, and (2) assigns layers from models to each time window. As the final assignment of layers to chiplets is not known apriori, the decisions in \textit{MCM-Reconfig} engine are based on expected execution times. Formally, given $|DF|$ dataflow style classes, the expected execution latency for a layer $l$ is:
\begin{equation}
\abovedisplayskip=5pt
    \mathbb{E}(Lat(l)) = \sum_{i=1}^{|DF|} \frac{n_{df_{i}}}{|\mathcal{C}|}\times Lat(l\rightarrow i) \label{eqn:expectation}
\belowdisplayskip=5pt
\end{equation}
where $n_{df_{i}}$ indicates the number of class $i$ chiplets integrated onto the MCM having $|C|$ chiplets in total; $Lat_{l \rightarrow i}$ is layer $l$ latency when scheduled on the class $i$ chiplet, which is retrieved offline from latency database generated by MAESTRO \cite{kwon2020maestro, kwon2019understanding}. The average execution time information is utilized in \textit{MCM-Reconfig} engine for window assignment process illustrated in~\autoref{fig:algorithm}.

\betterparagraph{Time Windows Characterization}
\textit{MCM-Reconfig} engine first specifies the number of windows, 
through a hyperparameter, $n_{splits}$, to explore proper cut points for each model. For example, in~\autoref{fig:algorithm}, the model A has a cut after layer 6, which led to having layers 1-6 in Window 1. The worst-case latency experienced by any model is set as the time horizon to be partitioned into periodic time windows. 

\betterparagraph{Greedy Layer Packing Algorithm}
We adopt a \textit{first-fit greedy-packing} heuristic to assign layers to execution time windows if their execution time is expected to finish within the time window boundaries (see Algorithm \ref{alg:layer_assign}).
Any layer whose execution time lies across two time windows is deferred to the next time window. 
Through this approach, we enable (i) running low-latency layers in earlier windows (restricts starvation). (ii) dynamically controlling the number of time windows by skipping trivial time windows with no workloads.

Based on our analysis of the periodic window characterization with greedy layer packing using a workload of UNet and GPT2-L against a layer-optimal approach. We found the rate of EDP improvement stagnated after 4 splits. We set $n_{splits}$=4 (5 time windows) as our default unless otherwise stated.

\begin{algorithm}[t]
\caption{Greedy Layer Packing Algorithm}\label{alg:layer_assign}
    \begin{algorithmic}[1]
        \footnotesize
        \Require{$M$ (workloads), $\mathcal{T}$, $\mathcal{C}$, $DF$}
        \Ensure{$L2W$ (Layer(s) to windows assignments)}
        \Function{LayerAssignment}{$M$, $\mathcal{C}$, $\mathcal{T}$}
            \For{$m \in M$}
                \State $exec\_win = ()$ 
                \State $win\_idx, used\_cycles = 0, 0$
                \For{$l \in m$}
                    \State $\mathbb{E}(Lat(l)) = \sum_{i=1}^{|DF|} \frac{n_{df_{i}}}{|\mathcal{C}|}\times Lat(l\rightarrow i)$
                    \While{True}
                        \If{$win\_idx == |\mathcal{T}|$} 
                        \State $Slack$ = $None$
                        \Else
                        \State $Slack$ = $\rho[win\_idx] - used\_cycles$ 
                        \EndIf
                        \If{$Slack$ == $None$ or $\mathbb{E}(Lat(l)) <= Slack$}
                        \State $exec\_win$ += $(l,)$
                        \State $used\_cycles$ += $\mathbb{E}(Lat(l))$
                        \State \textbf{Break}
                        \Else      
                        \State $L2W[win\_idx][m] = exec\_win$ 
                        \State $used\_cycles = \mathcal{T}[win\_idx]$
                        \State $exec\_win = ()$
                        \State $win\_idx$ += 1
                        \EndIf
                    \EndWhile
                \State $L2W[win\_idx][m] = exec\_win$ 
                \EndFor
            \EndFor
        \EndFunction
    \end{algorithmic}
\end{algorithm}

\subsection{Provisioner Engine \textit{(PROV)}}

The \textit{PROV} engine provides an initial estimate on the number of chiplet needed by each model workload in every time window from a candidate partitioning strategy. PROV assignments are agnostic to the underlying chiplets' properties (dataflow, resources), and hence we refer to chiplets in this state as \textit{nodes}.
We implement the \textit{PROV} engine to support exhaustive search or rule-based node distribution assignments. A uniform distribution rule allocates $N_i$ nodes to the $i^{th}$ model as follows:

\begin{equation}
    N_i = round(\frac{\mathbb{E}(\mathcal{P}_i)}{\sum_j(\mathbb{E}(\mathcal{P}_j)} \times |\mathcal{C}|)\label{eqn:prov}
\end{equation}
where $\mathbb{E}(\mathcal{P}_i)$ represents the expected value of a target performance optimization metric (latency, energy, EDP) for the model $i$. $\mathbb{E}(\mathcal{P}_i)$ is computed in a manner similar to the expectation formula in Equation (\ref{eqn:expectation}). 

We ensure every model in the time window is assigned at least one node to progress its execution. The rules enable trading off search complexity for coverage. We analyze the efficacy of the uniform distribution compared to the exhaustive search in \autoref{sec:eval}.

\subsection{Segmentation Engine \textit{(SEG)}} \label{subsec:segment}

The \textit{SEG} module is instantiated every time window to partition topologically sorted model layers into layer segments (\autoref{def:segment}) that are mappable to computing nodes for exclusive execution throughout the time window.
Different segmentation choices reflect various trade-off points between the \textit{layer-sqeuencing} and \textit{layer-pipelining} features: the former concerns with execution locality on the same node; the latter specifies inter-layer and -chiplet pipelining opportunities.

\textbf{Segmentation Search Space.} A segmentation candidate is represented by a sequence of splitting points. Candidate splitting points for a model can be specified after each layer provided to the \textit{SEG}.
Given $|L_i|$ and $|N_i|$ as the respective number of layers and number of assigned nodes from the \textit{PROV} to model workload $m_i$, the max number of segments that can be generated for $m_i$ is upper bounded by $N_i$. Thus, the overall segmentation space complexity becomes $\mathcal{O}(\Pi_i \ \binom{L_i}{N_i - 1})$, 
We incorporate the following heuristics to manage complexity.

\textbf{Heuristic 1. Product to summation reduction.}
We reduce complexity by leveraging the independence of segments from different models to divide the search into a two-step process: (1) \textit{SEG} first evaluates segmentation candidates from each model \textit{separately}. (2) The top-k segmentation candidates from each model are used to construct a smaller combinatorial segmentation search space used in the final evlauation. Through this heuristic, the search space complexity is reduced from $\mathcal{O}(\Pi_i \ \binom{L_i}{N_i - 1})$ to $\mathcal{O}(max(\binom{L_i}{N_i - 1}))$.

\textbf{Heuristic 2. Node allocation constraint.}
To further reduce the search complexity within the \textit{SEG} engine, SCAR supports having a node allocation constraint as a hyperparameter, particularly beneficial in cases where time windows have model workloads with dissimilar layer distributions (e.g., one model with a heavy layer, another with numerous small layers), which could lead to an explosion in the number of trivial segmentation options from the low-cost layers, causing the the \textit{SEG} search space complexity to rise. As such small layers are more suited for continuous execution on the same resource by virtue of their lower computational footprints, and to limit unnecessary, costly inter-chiplet data movements, we designate a node allocation constraint to enable alleviating this added complexity by restricting the number of nodes assigned to workloads with disproportionately large number of layers.

\begin{figure}[!tbp]
\begin{center}
{\includegraphics[,width = 0.44\textwidth]{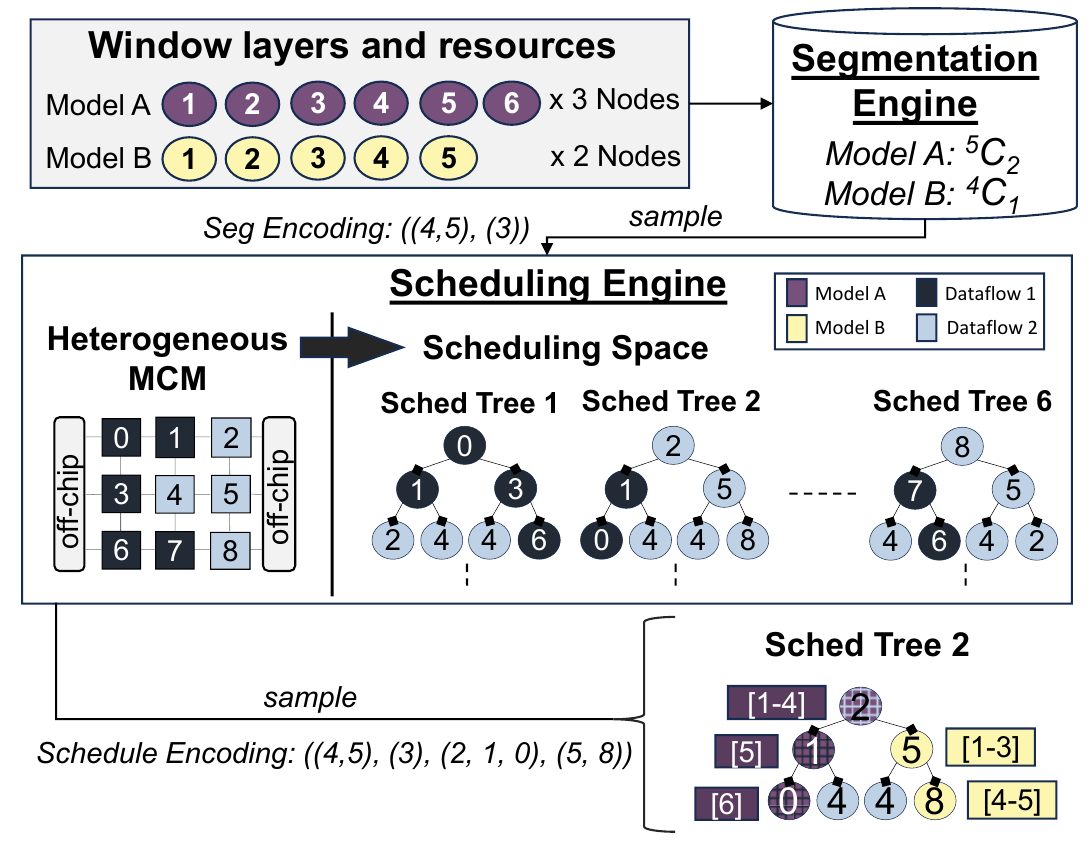}}
\end{center}
\vspace{-2.6ex}
\caption{Schedules creation through the \textit{SEG} and \textit{SCHED} engines.}
\label{fig:sched}
\vspace{-3.2ex}
\end{figure}

\subsection{Scheduling Engine (\textit{SCHED})}

The innermost \textit{SCHED} engine is responsible for generating the actual physical mapping of layer segments onto chiplets.

\betterparagraph{SCHED Search Space} As illustrated in Figure \ref{fig:sched}, the scheduling search space for the mapping of $M$ model workloads onto $\mathcal{C}$ chiplets is represented by a forest of scheduling trees. We define (i) \textit{forest}; as the entire collection of search trees. (ii) \textit{tree}; a single scheduling tree with all models. (iii) \textit{subtree}; a subset tree exclusively associated with a model. 

\betterparagraph{Scheduling Tree Composition} every node in a scheduling tree corresponds to a unique chiplet resource on the MCM showcasing its distinctive heterogeneous features (i.e., dataflow). 
Tree edges are constructed based on each chiplet's neighbors connected directly through an interposer. Though a node $j$ can be replicated throughout the tree, it can only be visited once, indicating its exclusive occupancy by a model.


\betterparagraph{Trees Distinction} within each tree, the root nodes of the subtrees specify different chiplets as potential starting positions for candidate model schedules (see Figure \ref{fig:sched}). 
Thus, the scheduling space coverage starts by selecting a tree, represented by a permutation sequence of subtrees' root nodes -- e.g., permutation sequence [i,j,k] indicates exploring scheduling candidates for a tree with scheduling candidates starting at chiplet positions $i$, $j$, and $k$ for a 3-model workload. The depth of model $i$'s subtree is determined by $N_i$.


\betterparagraph{Candidate Schedules Generation} Through traversing each subtree, we can obtain candidate execution schedules for each model by assigning segments orderly to the subtree's nodes.
Starting from the root node of the first model's subtree, a constrained depth first search (DFS) is performed generating a candidate schedule path once the full subtree depth ($N_i$) is reached. This traversal is repeated for each subsequent subtree, constrained on the preceding subtree's prior visited nodes. 

\betterparagraph{Encoding and Search Algorithm}
 As shown in
Figure \ref{fig:sched}, use a $2 \times |M|$-length tuple to represent the final scheduling encoding, where the first $M$ entries reflect segmentation decisions for each model $m_i$, and the latter $|M|$ entries reflect schedule mappings of segments to chiplets for each workload. This encoding enables various search strategies (e.g., evolutionary).


\betterparagraph{Search Space Complexity}
Given $|M|$ as the number of models in a given window, $|T|$ the number of scheduling trees in the search space, $d$ is a traversal path's degree of freedom, and $N_{max}$ representing the max number of resources allocated to any model in this window. The scheduling search complexity can be given by $\mathcal{O} (|M| \times |T| \times d^{N_{max}})$. 






\subsection{Cost Model and Scoring}
\label{subsec:cost_model}

\begin{table}[t]
    \centering
    \footnotesize
    \caption{MCM microarchitecture parameters from \cite{shao2019simba, orenes2023massive}. All numbers are scaled to 28 nm technology.}
    \vspace{-2ex}
    \begin{tabular}{l | c | c}
    \hline
    \multirow{3}{2pt}{Offchip Memory} & DRAM latency & 200 ns \\
    & DRAM energy & 14.8 pJ/bit \\
    & DRAM bandwidth & 64 GB/s \\
    \hline
    \multirow{3}{*}{Package} & NoP intercon. latency & 35 ns/hop \\
    & NoP intercon. energy & 2.04 pJ/bit \\
    & NoP intercon. bandwidth & 100 GB/s/Chiplet \\
    \hline
    
    \end{tabular}
    \label{tab:params}
    \vspace{-3ex}
\end{table}

We implement a cost model for evaluating scheduling candidates on different performance efficiency metrics. 

\betterparagraph{Cost Model} We build our analytical cost model on top of MAESTRO \cite{kwon2019understanding, kwon2020maestro}, which is a standardized analytical cost model tool for modeling performance of AI operators' mappings (dataflow and tiling) on AI accelerators, with a reported 96\% accuracy of performance estimation compared to low-level RTL simulation. MAESTRO natively enables \textit{intra-chiplet} performance modeling, and we extend it for the MCM AI accelerator setting by integrating the MCM design parameters from Simba \cite{shao2019simba} as shown in Table \ref{tab:params}, where we add MCM-specific layers upon MAESTRO leveraging its proven communication overhead modeling methodology to model \textit{offchip} and \textit{inter-chiplet} communication costs. The overall cost model follows the performance models in \autoref{subsec:lat}.


\betterparagraph{Scoring}
Scores are estimated based on latency, energy, or EDP metrics following the characterization in \autoref{sec:problem_formulation}. The \textit{SCHED} aggregates scores for each model's schedule, and returns the top performing configuration to the \textit{SEG} engine to rank segmentation strategies. Top segmentation strategies in each window are aggregated to score the overall scheduling strategy at \textit{MCM-Reconfig} (see the scoring flow in Figure \ref{fig:framework}). 

\section{Evaluation}
\label{sec:eval}

\subsection{Experimental Settings} \label{subsec:settings}

\begin{table}[t]
\scriptsize
    \centering
    \caption{Our experimental Multi-model workload scenarios for datacenter and AR/VR use-cases inspired by MLPerf \cite{reddi2020mlperf, mlcommons} and XRbench \cite{kwon2023xrbench} benchmarks. `sl' indicates sequence length} 
    
    
    \vspace{-2ex}
    \begin{tabular}{ c | c | c | c}
    \hline
    Use-Case & Scenario & Models & Batch Size \\
    \hline
    \multirow{18}{6em}{{Datacenter (MLPerf) \cite{reddi2020mlperf, mlcommons}}} & \multirow{2}{6em}{{(1) LMs}} & GPT-L \cite{radford2019language} (sl=128) & 1\\ 
    & & BERT-L \cite{devlin2018bert} (sl=128)  & 3\\
    \cline{2-4}
    & \multirow{3}{6em}{{(2) LMs + Image}} & GPT-L \cite{radford2019language} (sl=128) & 1\\ 
    & & BERT-L \cite{devlin2018bert} (sl=128)  & 3\\
    & & ResNet-50 \cite{he2015deep} (224$\times$224$\times$3) & 1\\
    \cline{2-4}
    & \multirow{3}{6em}{{(3) LMs + Image}} & GPT-L \cite{radford2019language} (sl=128) & 1\\ 
    & & BERT-L \cite{devlin2018bert} (sl=128)  & 3\\
    & & ResNet-50 \cite{he2015deep} (224$\times$224$\times$3) & 32\\
    \cline{2-4}
    & \multirow{4}{6em}{{(4) LMs + Segmentation + Image}} & GPT-L \cite{radford2019language} (sl=128) & 8\\ 
    & & BERT-L \cite{devlin2018bert} (sl=128)  & 24\\
    & & U-Net \cite{unet} (512$\times$512$\times$1) & 1 \\
    & & ResNet-50 \cite{he2015deep} (224$\times$224$\times$3) & 32\\
    \cline{2-4}
    & \multirow{6}{6em}{{(5) LMs + Segmentation + Image}} & GPT-L \cite{radford2019language} (sl=128) & 8\\ 
    & & BERT-L \cite{devlin2018bert} (sl=128)  & 24\\
    & & BERT-base \cite{devlin2018bert} (sl=128) & 24\\
    & & U-Net \cite{unet} (512$\times$512$\times$1) & 1 \\
    & & ResNet-50 \cite{he2015deep} (224$\times$224$\times$3) & 32\\
    & & GoogleNet \cite{szegedy2015going} (224$\times$224$\times$3) & 32\\
    \hline\hline

\multirow{15}{6em}{{AR/VR (XRBench) \cite{kwon2023xrbench}}} & \multirow{5}{6em}{{(6) AR Assistant}} & D2GO \cite{D2go} (Object Det.) & 10\\ 
    & & PlaneRCNN \cite{liu2019planercnn} (Plane Det.) & 15\\
    & & MiDaS \cite{ranftl2020towards} (Depth Est.) & 30\\
    & & Emformer \cite{shi2021emformer} (Speech Rec.) & 3\\
    & & HRViT \cite{hrvit} (Semantic Seg.) & 10\\
    \cline{2-4}
    & \multirow{3}{6em}{{(7) AR Gaming}} & PlaneRCNN \cite{liu2019planercnn} (Plane Det.) & 15\\ 
    & & Hand S/P \cite{ge20193d} (Hand Track.)  & 45\\
    & & MiDaS \cite{ranftl2020towards} (Depth Est.) & 30\\
    \cline{2-4}
    & \multirow{2}{6em}{{(8) Outdoors}} & D2GO \cite{D2go} (Object Det.) & 30\\ 
    & & Emformer \cite{shi2021emformer} (Speech Rec.) & 3\\
    \cline{2-4}
    & \multirow{3}{6em}{{(9) Social}} & EyeCod \cite{you2022eyecod} (Gaze Est.) & 60\\ 
    & & Hand S/P \cite{ge20193d} (Hand Track.) & 30\\
    & & Sp2Dense \cite{ma2018sparse} (Depth Ref.) & 30 \\
    \cline{2-4}
    & \multirow{2}{6em}{{(10) VR Gaming}} & EyeCod \cite{you2022eyecod} (Gaze Est.) & 60\\ 
    & & Hand S/P \cite{ge20193d} (Hand Track.)  & 45\\
    \hline\hline

    \end{tabular}
    \label{tab:scenarios}
    \vspace{-1ex}
\end{table}

\begin{figure}[!tbp]
\begin{center}
{\includegraphics[,width = 0.46\textwidth]{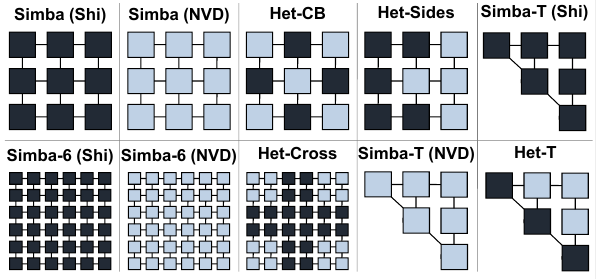}}
\end{center}
\vspace{-1ex}
\caption{Evaluated MCM chiplet organizations in this work.}
\label{fig:pattrn}
\vspace{-2ex}
\end{figure}

\betterparagraph{Multi-Model Workloads}
Our evaluations are performed on multi-model workload scenarios based on models from (1) MLPerf inference benchmark \cite{reddi2020mlperf, mlcommons}, which are curated for datacenter multi-tenancy setting following data
center usage trends in \cite{hazelwood2018applied, park2018deep, jouppi2017datacenter}. (2) XRBench \cite{kwon2023xrbench} for multi-model AR/VR workloads. The full list of scenarios is provided in Table \ref{tab:scenarios} covering a wide range of use-cases with varying degrees of diversity and complexity. 

\betterparagraph{MCM System} We evaluate SCAR on a variety of MCM systems following Simba architecture \cite{shao2019simba}. Simba comprises a total of 36 chiplets arranged as four $3\times3$ groups of chiplets and connected in a Mesh topology. We implement (1) $3\times3$, and (2) $6\times6$ MCM templates for our experiments. we adopt XY routing for on-pacakge data movement, and integrate further memory interfaces on the sides of the outer chiplets for direct access to the offchip DRAM as in \cite{gao2019tangram}.
We consider 4096 PEs/chiplet and 256 PEs/chiplet for the datacenter and AR/VR settings, respectively. We set L2 shared memory size in each chiplet to 10 MB as in  a recent mobile accelerator's on-chip memory size~\cite{qualcomm_hexagon_680}.

\betterparagraph{Baselines and MCM patterns}
We choose Shidiannao \cite{du2015shidiannao} and NVDLA \cite{nvdla} dataflow styles for our accelerator chiplets given their proven superiority \cite{kwon2021heterogeneous} and adopt two baselines:
\begin{itemize}
    \item \textit{Standalone}. each model is assigned a single accelerator chiplet - all chiplets adopt the same dataflow.
    \item \textit{Simba-like Pipelining}. In each time window, model workloads can be assigned to more than one chiplet. All chiplets adopt the same dataflow.
\end{itemize}

Figure \ref{fig:pattrn} illustrates the MCM chiplet patterns evaluated through SCAR. The 3$\times$3 MCM patterns are the default.

\betterparagraph{Optimization Targets} We perform our search space exploration experiments to target optimizing a single metric at a time, coining the terms Latency Search, Energy Search, and EDP Search. EDP Search is our default experiment.

\betterparagraph{Search Criteria}
We adopt an exhaustive brute-force search for all $3\times3$ MCM experiments. For the $6\times6$ experiment, we implement an evolutionary algorithm for the SEG module to navigate the rising complexity. We set the population size and max number of generations to 10 and 4, respectively.

\begin{table*}[ht]
    \setlength{\tabcolsep}{3pt}
    \centering
    \caption{Breakdown of the search results across all MLPerf datacenter scenarios from Table \ref{tab:scenarios}. The comparison involves the top performing models on latency and EDP for each search strategy across all workload scenarios. Latency estinates at 500 MHz. }


    \vspace{-2ex}
    \begin{tabular}{| l | c c c c c | c c c c c | c c c c c | c c c c c |}
    \hline
    

    \multirow{3}{*}{Strategy} & \multicolumn{10}{c|}{Latency Search} & \multicolumn{10}{c|}{EDP Search} \\ \cline{2-21}
    & \multicolumn{5}{c|}{Latency (s)} & \multicolumn{5}{c|}{EDP (J.s)} & \multicolumn{5}{c|}{Latency (s)} & \multicolumn{5}{c|}{EDP (J.s)} \\\cline{2-21}
    & Sc1 & Sc2 & Sc3 & Sc4 & Sc5 & Sc1 & Sc2 & Sc3 & Sc4 & Sc5 & Sc1 & Sc2 & Sc3 & Sc4 & Sc5 & Sc1 & Sc2 & Sc3 & Sc4 & Sc5 \\\cline{1-21}
    Stand.(Shi) & 1.41 & 1.41 & 2.63 & 14.82 & 14.82 & 0.029 & 0.047 & 1.21 & 16.43 & 21.09 & 1.41 & 1.41 & 2.63 & 14.82 & 14.82 & 0.029 & 0.047 & 1.21  & 16.43 & 21.0 \\
    Stand.(NVD) & 0.36 & 0.36 & 4.06 & 7.94 & 7.94 & 0.007 & 0.01 & 0.786  & 5.707 & 7.608 & 0.36 & 0.36 & 4.06 & 7.94 & 7.94 & \textbf{0.007} & \textbf{0.01} & 0.786  & 5.707 & 7.608 \\
    Simba (Shi) & 0.99 & 0.99 & 2.44 & 7.98 & 8.29 & 0.02 & 0.03 & 1.048 & 10.85 & 11.216 & 1.13 & 0.97 & 2.53 & 8.03 & 9.79 & 0.024 & 0.024 & 1.097 & 8.706 & 13.75 \\
    Simba (NVD) & \textbf{0.25}  & \textbf{0.27} & 1.37 & 8.34 & 8.34 & \textbf{0.005} & \textbf{0.007} & \textbf{0.24} & 6.216 & 8.225 & \textbf{0.29}  & \textbf{0.34} & 1.36 & 8.284 & 8.28 & 0.006 & 0.008  & \textbf{0.238} & 6.165 & 8.162 \\
    Het-CB & 0.36 & 0.39  & \textbf{1.02} & 6.04 & 7.39 & 0.007 & 0.013 & 0.331 & 5.545 & 9.051 & 0.51 & 0.39 & \textbf{1.27} & 5.6 & 7.81 & 0.01 & 0.013 & 0.37 & 5.21 & 9.166\\
    Het-Sides & 0.34  & 0.34 & 1.12 & \textbf{3.07} & \textbf{4.81} & 0.007 & .011 & 0.463 & \textbf{3.074} & \textbf{5.8} & 0.36  & 0.38 & 2.0 & \textbf{3.8} & \textbf{5.19} & 0.008 & 0.01 & 0.35 & \textbf{3.328} & \textbf{6.107}\\ \hline\hline



    \end{tabular}
    \label{tab:breakdown}
    \vspace{-2.5ex}
\end{table*}

\subsection{Datacenter multi-model scheduling results}

\begin{figure}[!tbp]
\begin{center}
{\includegraphics[,width = 0.48\textwidth]{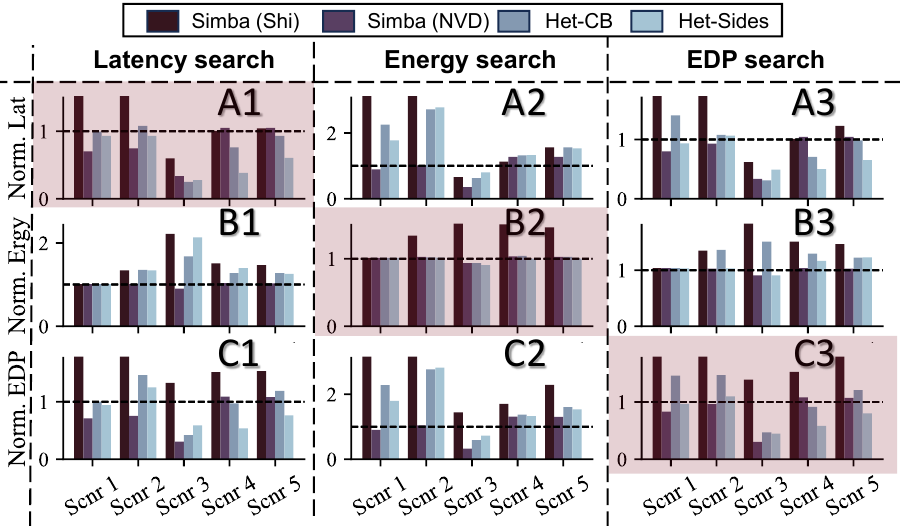}}
\end{center}
\vspace{-3ex}
    \caption{Latency, energy, and EDP evaluations for top-scoring 3$\times$3 candidates normalized by the standalone NVDLA for datacenter scenarios. Highlighted barplots indicate main results (aligned optimization and evaluation metric).
    }
\label{fig:barplots}
\vspace{-3ex}
\end{figure}

\begin{figure}[!tbp]
\begin{center}
{\includegraphics[,width = 0.48\textwidth]{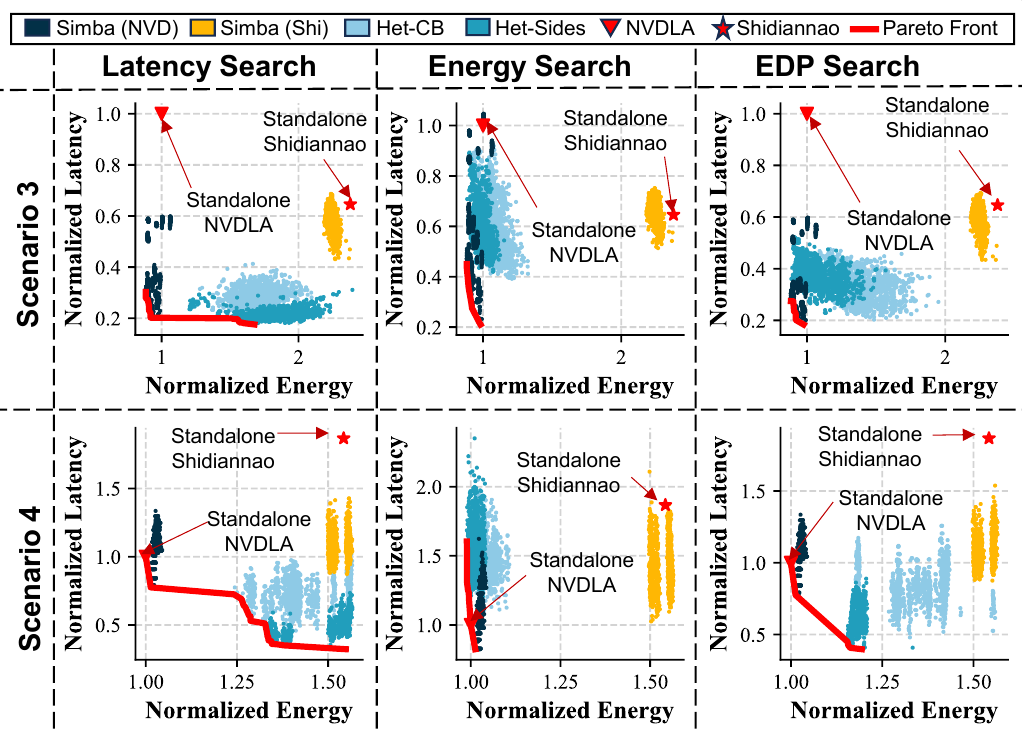}}
\end{center}
\vspace{-2.6ex}
    \caption{Pareto results for the brute force search across various MCM strategies on various search targets for scenarios 4 and 5 from Table \ref{tab:scenarios}. }
\label{fig:pareto}
\vspace{-2.5ex}
\end{figure}

\begin{figure}[!tbp]
\begin{center}
{\includegraphics[,width = 0.49\textwidth]{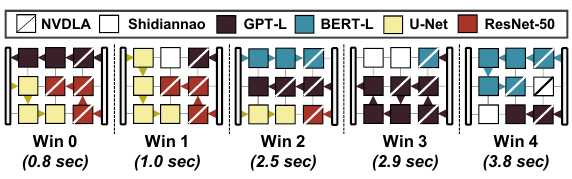}}
\end{center}
\vspace{-3ex}
    \caption{Top-scoring scheduling strategy for Scenario 4 Het-Sides. Allocation of chiplets and coarse-grained schedules are shown for each time window. Times computed over 500 MHz indicate cumulative window latencies.}
\label{fig:top_cand}
\vspace{-3ex}
\end{figure}

We discuss the full datacenter scheduling results on the $3\times3$ MCM, provided in Table \ref{tab:breakdown} and Figure \ref{fig:barplots}. Figure \ref{fig:pareto} illustrates the Pareto search results for Scenarios 3 and 4.

\betterparagraph{Impact of Workload Scenarios} We observe that different scenarios are more affine towards different MCM strategies owing to their varying degrees of diversity and computational demands within the multi-model suite. For example in the EDP search in Table \ref{tab:breakdown}, scenarios 1-3 favor the standalone (NVD) and Simba (NVD) strategies as both are dominated by low-batched, transformer-based workloads (GPT-L and BERT-L) characterized by strong affinities towards the NVDLA dataflow style, As the computational load and model diversity increase, schedules obtained through heterogeneous strategies become more favorable. 
For instance in the EDP search scenarios 4-5 with more batches and heavy ResNet-50 and UNet workloads, Het-Sides outperforms all other strategies on EDP evaluation, experiencing 46.02\% and 25.18\% less EDP compared to Simba (NVD) on the respective scenarios.

\betterparagraph{Target Objective Impact} In matching criteria plots (A1, B2, and C3) from \autoref{fig:barplots}, Het-Sides schedules for the heaviest scenarios 4-5 outperform schedules from all other strategies. This superiority, however, comes at the expense of other metrics. For instance in C3, the Het-Sides schedule outperforms that from all other strategies by experiencing 0.58$\times$ the EDP from the Standalone (NVD). This is achieved, however, by speeding up execution at the expense of energy consumption, where Het-Sides achieves the fastest execution speedup of 2$\times$ over Standalone (NVD) in (A3) at the expense of being $1.22\times$ more energy demanding than Standalone (NVD) in (B3).

\betterparagraph{Pipelining Benefits and NoP impact} 
Pipelining individually can offer performance improvements over standalone baselines when ample resources are available to the models. For example, Scenario 3 (Latency Search) -- top-left most Pareto in \autoref{fig:pareto} -- shows Simba (NVD) schedules achieving considerable speedups over the the standalone NVDLA baseline (2.9 $\times$ for the top-performing candidate from Table \ref{tab:breakdown}). The reason being that as the scheduler targets latency optimization, it attempts to consistently identify chiplet assignment strategies that equate pipelining latencies between model workloads assigned to the same time window. For instance, in the last time window containing ResNet-50, the Simba (NVD) scheduler for Scenario 3 queued 18, 14, and 29 layers from the GPT-L, BERT-L, and ResNet-50 for processing, provisioned 3 chiplets for each subset, and yielded accordingly 3 layer segments per model. This configuration led to the model workloads achieving comparable inter-chiplet pipelining latencies of 0.28 ms, 0.32, and 0.33 ms, eventually leading the full schedule to sustain a total of 1.37 s latency compared to the 4.06 s from the standalone NVDLA. We also observe that keeping medium-sized data traffic on package through chiplet-to-chiplet data passing contributes additional energy savings for these medium sized workloads by decreasing offchip memory read/write accesses at the segments' start and end times.

For the NoP traffic, \autoref{fig:barplots}-A1 summarizes our findings. In Scenarios 1-3, Simba (NVD) leverages inter-chiplet pipelining to outweigh added NoP costs, achieving latency speedups over standalone NVDLA reaching $1.4\times$, $1.3\times$, and $2.9\times$, respectively. In Scenarios 4-5, Simba (NVD) to incur $1.05\times$ slowdown from the added traffic in both scenarios.

\betterparagraph{Heterogeneity Benefits} Heterogeneous integration patterns compensate for the added NoP traffic in the heavier workloads (Scenarios 4-5) and boost efficiency through considering per-layer dataflow affinities. Het-Sides achieves 1.7$\times$ and $1.25\times$ EDP efficiency over the standalone (NVD). We find also that Het-Sides is always superior to Het-CB in such heavy scenarios as it offers opportunities for both homogeneous and heterogeneous inter-chiplet pipelining. This is beneficial for batched layer sequences with same dataflow affinities.

\betterparagraph{Het-Sides Top Schedule}
In Figure \ref{fig:top_cand}, we illustrate the top-scoring Het-Sides schedule from the EDP search in Scenario 4. 
As shown, the greedy-packing algorithm leads to non-uniform windows where smaller workloads (ResNet-50) are assigned to the earlier windows at the expense of larger workloads (BERT-L).
This facilitates (i) fine grained optimization of small workload schedules; (ii) avoiding small workloads starvation. GPT-L and BERT workloads dominate the schedule from window 2. 
\autoref{tab:top_lat} breaks down the latency in each window. 

\subsection{AR/VR multi-model scheduling results}

\begin{figure}[!tbp]
\begin{center}
{\includegraphics[,width = 0.48\textwidth]{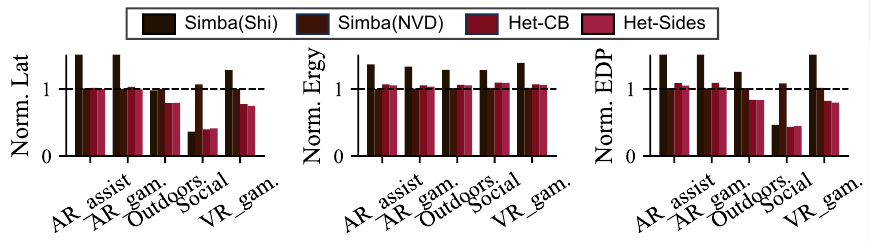}}
\end{center}
\vspace{-2.8ex}
    \caption{Evaluations on the EDP search for the XRBench usage scenarios listed in Table \ref{def:workload_scenario} normalized by NVDLA standalone configuration.}
\label{fig:vr_barplot}
\vspace{-2.2ex}
\end{figure}

\begin{figure}[!tbp]
\begin{center}
{\includegraphics[,width = 0.48\textwidth]{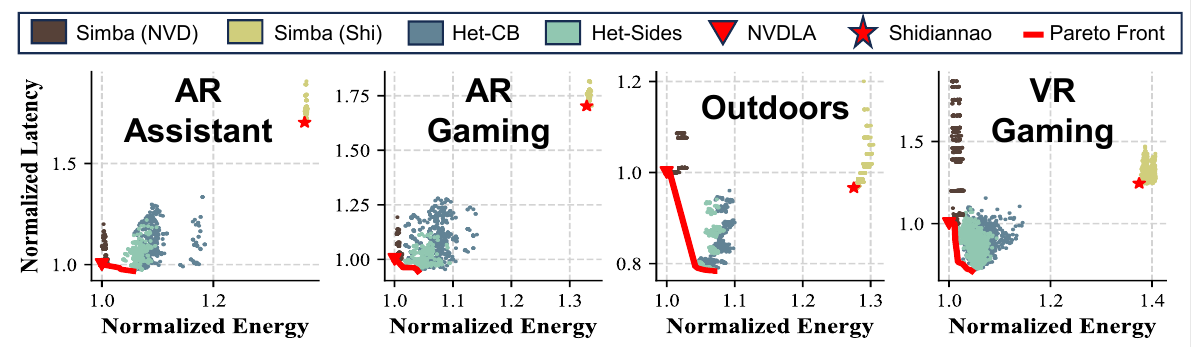}}
\end{center}
\vspace{-2.5ex}
    \caption{Pareto optimal results on the EDP search experiments for the labeled XRBench usage scenarios. Results normalized by standalone NVDLA}
\label{fig:vr_pareto}
\vspace{-2ex}
\end{figure}

The AR/VR scheduling results on the EDP search for the $3\times3$ MCM are provided in Table \ref{tab:vr_breakdown} and Figure \ref{fig:vr_barplot}. Figure \ref{fig:vr_pareto} illustrates their the Pareto search results. On average, the Het-Sides schedule achieves $17\%$ and $21.3\%$ EDP improvement over standalone NVDLA and Simba (NVD), respectively. We also find that the smaller MCM system limits improvement potential for the heaviest scenarios, AR assistant (scenario 6) and AR gaming (scenario 7) due to increased resource contention, leading the heterogeneous and NVDLA based schedules to achieve comparable evaluations (see \autoref{fig:vr_barplot}).

\begin{table}[t]
    \setlength{\tabcolsep}{3pt}
    \scriptsize
    \centering
    \caption{EDP search AR/VR results normalized by standalone NVDLA}
    \vspace{-2ex}
    \begin{tabular}{| l | c c c c c | c c c c c |}
    \hline
    \multirow{3}{*}{Strgy} & \multicolumn{10}{c|}{EDP Search} \\ \cline{2-11}
    & \multicolumn{5}{c|}{Relative Latency} & \multicolumn{5}{c|}{Relative EDP} \\ \cline{2-11}
    & Sc6 & Sc7 & Sc8 & Sc9 & Sc10 & Sc6 & Sc7 & Sc8 & Sc9 & Sc10 \\ \hline
    Stand.(Shi) & 1.7 & 1.7 & 0.96 & 0.36 & 1.24 & 2.32 & 2.26 & 1.22 & 0.45 & 1.71 \\
    Stand.(NVD) & 1.0 & 1.0 & 1.0 & 1.0 & 1.0 & 1.0 & 1.0 & 1.0 & 1.0 & 1.0 \\
    Simba(Shi) & 1.72 & 1.75 & 0.97 & 0.36 & 1.28 & 2.36 & 2.33 & 1.24 & 0.47 & 1.78 \\
    Simba(NVD) & 1.01 & 1.0 & 0.99 & 1.07 & 1.0 & 1.01 & 1.01 & 1.0 & 1.08 & 1.02 \\
    Het-CB & 1.01 & 1.03 & 0.78 & 0.4 & 0.77 & 1.09 & 1.09 & 0.83 & 0.43 & 0.82 \\
    Het-Sides & 1.0 & 0.99 & 0.78 & 0.41 & 0.75 & 1.05 & 1.03 & 0.83 & 0.45 & 0.79 \\ \hline\hline
    \end{tabular}
    \label{tab:vr_breakdown}
    \vspace{-2.5ex}
\end{table}

\begin{table}[t]
    \scriptsize
    \centering
    \caption{End-to-end latency breakdown in seconds for the top partitioning candidate in Figure \ref{fig:top_cand}. }
    \vspace{-2ex}
    \begin{tabular}{ l | c  c  c  c  c |c  c| c }
    \hline
     & \textbf{W0} & \textbf{W1} & \textbf{W2} & \textbf{W3} & \textbf{W4} & \textbf{ideal} & \textbf{tot} & \#\textbf{layers} \\
    \hline
    \textbf{GPT-L} & 0.23 & 0.21 & 1.02 & 0.28 & 0.23 & 1.97 & 3.1 & 120 \\ 
    \textbf{BERT-L} & 0 & 0 & 1.47 & 0.4 & 0.90 & 2.77 & 3.76 & 60 \\
    \textbf{U-Net} & 0.21 & 0.14 & 0.46 & 0 & 0 & 0.8 & 1.45 & 23 \\
    \textbf{ResNet} & 0.78 & 0.17 & 0.11 & 0 & 0 & 1.1 & 1.1 & 66 \\
    \hline
    \textbf{Window} & 0.78 & 0.21 & 1.47 & 0.4 & 0.9 & - &  3.8 & \multirow{2}{*}{269}\\ 
    \cline{1-8}
    \textbf{$\#$layers} & 60 & 30 & 131 & 25 & 23 & - & - & \\ 
    \hline
    \end{tabular}
    \label{tab:top_lat}
\end{table}

\begin{figure}[!tbp]
\begin{center}
{\includegraphics[,width = 0.48\textwidth]{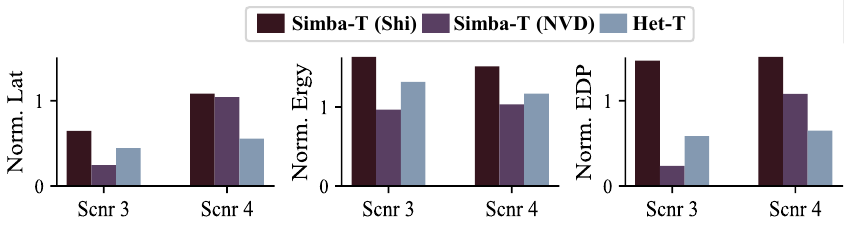}}
\end{center}
\vspace{-3ex}
    \caption{Results for the EDP Search for Scenarios 3 and 4 on the Triangular NoP topologies from Figure \ref{fig:pattrn} (Normalized by standalone NVDLA).}

\label{fig:ablation_2x3}
\vspace{-3ex}
\end{figure}

\subsection{Scaling to 6$\times$6 MCM system}
\begin{figure}[!tbp]
\begin{center}
{\includegraphics[,width = 0.48\textwidth]{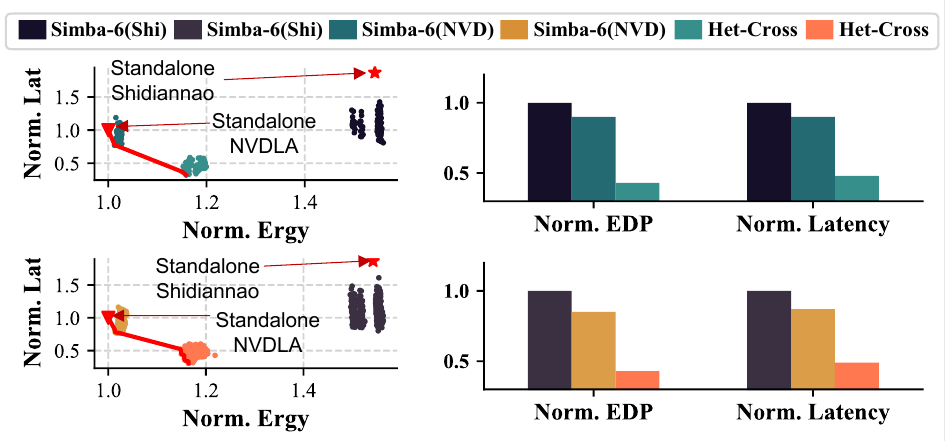}}
\end{center}
\vspace{-3ex}
    \caption{Pareto Plot and comparison of top performing models for the EDP search on the $6\times6$ MCM at $n_{splits}$=2 (\textit{top}) and $n_{splits}$=3 (\textit{bot}). }

\label{fig:ea_barplot}
\vspace{-3.5ex}
\end{figure}

We assess how SCAR scheduling scales with the hardware using the 6$\times$6 full Simba MCM system with its templates from Figure \ref{fig:pattrn}. We implement an evolutionary algorithm for the \textit{SEG} engine to scale with the rising problem size, and show the results for the EDP search on Scenario 4 at $n_{splits}$=$2$ and $n_{splits}$=$3$. We choose the heterogeneous cross (Het-Cross) as our heterogeneous template following the insights from the 3$\times$3 experiment regarding choosing heterogeneity patterns that enable both homogeneous and heterogeneous pipelining capabilities. As shown in Figure \ref{fig:ea_barplot}, the evolutionary search has led to identifying configurations for Het-Cross that achieve 2.3$\times$ and 1.9$\times$ reduction in EDP; 2.1$\times$ and 1.8$\times$ reduction in latency over Simba (Shi) and Simba (NVD), respectively.

\subsection{Ablation Studies}
\betterparagraph{Ablation Study on Time Partitioning}Using Scenario 4 and Het-Sides strategy, we study how performance changes when varying $n_{splits}$ from 1 to 5 and repeating the EDP search experiment. Prior to $n_{splits}$=4, the average rate of reduction in EDP was 1.25$\times$. The rate of improvement drops to only 1.04$\times$ between $n_{splits}$=4 to 5, indicating diminishing returns.

\betterparagraph{Ablation Study on other NoP Topology} SCAR can generalize to other NoP topologies as it relies on adjacency matrix connectivity. We test this by performing the EDP search for Scenarios 3 and 4 using the triangular NoP topologies in Figure \ref{fig:pattrn}. As shown in Figure \ref{fig:ablation_2x3}, similar performance patterns are exhibited compared to the Scenario results on the 3$\times$3 Mesh, albeit with varying relative gains. For instance, though Het-T outperforms both Simba-T strategies in Scenario 4 by 2.5$\times$ and 1.67$\times$ over Simba-T (Shi) and Simba-T (NVD), respectively, it is second best compared to Simba-T (NVD) in Scenario 3 by a factor of 2.5$\times$ (compared to 1.47$\times$ from \autoref{fig:barplots}-C3) due to the increased competition for resources.

\betterparagraph{{Ablation Study on rule-based PROV}} We repeated the EDP search for all strategies across Scenarios 3-5 on the $3\times3$ template using an exhaustive search over the $N_i$ values. Though the results improved further with the added search complexity, the insights remained the same. For Scenarios 4-5, Het-Sides remained superior, achieving respective EDP reductions of 38.3\% and 29.9\% from Standalone NVDLA; 59.5\% and 57.6\% from Simba (Shi); 33.0\% and 28.3\% from Simba (NVD). For Scenario 3, Simba (NVD) remained superior, achieving 79.7\% EDP reduction Het-Sides. These evaluations follow the trends from our rule-based results in Table \ref{tab:breakdown}.

\betterparagraph{Ablation on Greedy Packing Algorithm} Using Scenario 4 and Het-Sides, we test the efficacy of our first-fit greedy layer packing algorithm against a uniform packing baseline, distributing layers uniformly across time windows. Ours achieved 21.8\% speedup and 8.6\% energy reduction.

\subsection{Summary of Results and Main Insights}

We summarize our main insights and findings as follows.

\begin{itemize}
    \item Heterogeneous MCM patterns improve performance for heavy and diverse multi-model workloads (scenarios 4-5).
    \item Homogeneous MCM patterns are more suited for small multi-model workloads (scenarios 1-3).
    \item Heterogeneous MCM patterns with diverse pipelining options (Het-Sides) are superior to heterogeneous patterns with homogeneous pipelining options (Het-CB). 
    \item The target optimization objective is crucial in identifying the best integration strategy. In EDP search scenario 4, Het-sides outperformed all other strategies on EDP, but not on pure energy consumption.
    \item Topology and number of resources affect the extent of performance improvement for heterogeneous strategies.
\end{itemize}

Our findings show that understanding multi-model workload characteristics and usage scenarios is crucial for identifying the best MCM integration strategy for a target objective.
\section{Discussion and Limitations}
\label{sec:discussion}

\textbf{Multi-model optimization targets.} We experimented with different optimization targets (latency, EDP, energy) for our scenarios, and showed that the top performing strategy can change based on the target objective. As multi-model workloads evolve, it may be desirable to assign separate optimization targets for different models within a scenario (EDP v. lat). One practical way to achieve this in our framework is by adding a constraint in our EDP search, invalidating schedules that have certain models violate a latency constraint (i.e., the EDP search becomes lower bounded by the latency search).

\textbf{Heterogeneous chiplets technology.} Heterogeneous chiplet integration has become a viable, cost-effective approach to design state-of-the-art AI systems. Nvidia's world-class superchips are a successful example of heterogeneous on-package integration (e.g., Grace-Blackwell (1 CPU + 2 GPUs) \cite{nvidia_superchip}). The success of these systems and others (AMD's MI300X \cite{amdmi300x}) is testament to the hardware manufacturers' investment in chiplets technology, where through advanced manufacturing processes and heterogeneous integration capabilities, the development of MCM AI accelerators (like Nvidia's Simba \cite{shao2019simba}) becomes more accessible, allowing chiplet modifications/replacement in MCM hardware at lower costs without requiring a complete overhaul of the entire package.

\textbf{Scheduler Software Integration.} SCAR can be integrated on top of existing compiler infrastructure. The advanced scheduling techniques supported by the scheduler (dynamic chiplet regrouping, inter-chiplet pipelining) represent high-level abstractions of the computational graphs that can be transformed through standard compiler software (e.g., MLIR \cite{lattner2021mlir}) to representations suited for the underlying hardware. For example, dynamic chiplets regrouping is correspondent to graph partitioning, where a model's computational graph is divided into smaller subgraphs, each associated with the set of computing nodes assigned during the corresponding time window. The subgraphs can then be transformed to lower representations covering the details of buffer management, die-to-die communication, memory R/W requests, I/O, all the way to the transformations covering the dataflow features (loop reordering, spatial unrolling) for the specialized accelerators.

\section{Related Works}
\label{sec:related_works}

\begin{table}[t]
\scriptsize
    \centering
    \caption{Comparison against prior related scheduling works.}
    \vspace{-2ex}
    \begin{tabular}{ l | c c c c }
    \hline
     \multirow{2}{*}{\textbf{Work}} & \textbf{Chiplet-based} & \textbf{Multi-} & \textbf{Inter-Layer} & \textbf{Heterog-}\\
      & \textbf{{Systems}} & \textbf{Models} &  \textbf{{Pipelining}} & \textbf{{Aware}}\\
     \hline
     Simba\cite{shao2019simba} & \textbf{\checkmark} &  & \checkmark &\\
     Tangram\cite{gao2019tangram} & & & \checkmark &\\
     NN-baton\cite{tan2021nn}& \textbf{\checkmark}  & & &\\
     SET\cite{cai2023inter} & &  & \checkmark & \\
     Gemini\cite{cai2024gemini} & \checkmark &  & \checkmark & \\
    \hline
    Herald\cite{kwon2021heterogeneous}  & & \checkmark & & \checkmark \\
    MAGMA\cite{kao2022magma} &  & \checkmark & & \checkmark \\
    Planaria\cite{ghodrati2020planaria}  & & \checkmark & & \checkmark\\
    Veltair\cite{liu2022veltair}   & & \checkmark & & \\
    MoCA \cite{kim2023moca}  & & \checkmark & & \\
    \hline
    \textbf{This Work} & \checkmark & \checkmark & \checkmark & \checkmark \\
    \hline
    \end{tabular}
    \label{tab:realted}
    \vspace{-3ex}
\end{table}

\betterparagraph{Scheduler for Accelerators} Table \ref{tab:realted} compares our work against prior scheduling works. As shown, the related works can be categorized into two groups: one which has considered aspects of inter-layer pipelining and chiplet-based systems \cite{shao2019simba,gao2019tangram,tan2021nn, cai2023inter, cai2024gemini}, and another that focused on multi-model workloads on heterogeneous platforms \cite{kwon2021heterogeneous, kao2022magma, ghodrati2020planaria, liu2022veltair, kim2023moca}. Only this work addressed MCM, multi-model workloads, inter-layer pipelining, and heterogeneous dataflow. 

\betterparagraph{Multi-chiplet Modules} Several works proposed to address the scalability challenge for DNN acceleration via MCM integration \cite{shao2019simba, hwang2020centaur, orenes2023massive, tan2021nn, arunkumar2017mcm}. Most notably, Simba \cite{shao2019simba} pioneered a scalable deep learning MCM inference accelerator leveraging non-uniform work partitioning, communication-aware data placement, and cross-layer pipelining. 


\betterparagraph{Intra- and Inter-layer Parallelism} 
Prior works explored intra-layer parallelism to maximize DNN performance efficiency by partitioning DNN layers into smaller, parallelizable tiles \cite{parashar2019timeloop, hegde2021mind, huang2021cosa, lu2021tenet, xiao2021hasco, wu2022sparseloop, hong2023dosa}. Other works studied the inter-layer scheduling space to compensate for workloads with low degrees of parallelism \cite{gao2019tangram, kao2020confuciux, ma2021nasa, zheng2022atomic, cai2023inter, odema2023magnas, bouzidi2023map}.

\section{Conclusion}
\label{sec:conclusion}
In this work, we explored the scheduling space of a new class of MCM accelerator architecture, heterogeneous MCM AI accelerator, targeting multi-model AI workloads. We identify that the scheduling problem is intractably large but multi-level problem formulation and heuristics we proposed are effective for the large-scale scheduling problem. The results also show that heterogeneous MCM accelerator is beneficial for multi-model workloads, which motivates further exploration.


\bibliographystyle{plain}
\bibliography{ref}


\end{document}